\def\nll{\smallskip\hfil\hfil\linebreak\noindent} % New line with no indent: 2 \hfil's are necessary
\def\hi#1#2{$#1$\kern -2pt-#2} % hyphen \hi{N}{body} = N-body 
\def\hy#1#2{#1-\kern -2pt$#2$} % hyphen hy{large}{N} = large-N 
  
\documentclass[ams]{revtex4}
\usepackage{graphicx}
\begin{document}
% -----------------------------------------------
\hspace*{5.4 in}CUQM-128
\vspace*{0.4 in}
% -----------------------------------------------
\title{Comparison theorems for the Klein-Gordon equation in $d$ dimensions}
\author{Richard L. Hall and M. D. S. Aliyu}
\address{Department of Mathematics and Statistics, Concordia University,
1455 de Maisonneuve Boulevard West, Montr\'eal,
Qu\'ebec, Canada H3G 1M8}

\email{rhall@mathstat.concordia.ca}
\begin{abstract}
Two comparison theorems are established for discrete eigenvalues of the Klein-Gordon equation with an attractive central vector potential in $d\ge 1$ dimensions.~~(I)~If $\psi_1$ and $\psi_2$ are node-free ground states corresponding to positive energies $E_1\ge 0$ and $E_2\ge 0,$ and $V_1(r)\le V_2(r)\le 0$, then it follows that $E_1 \le E_2.$~~(II)~If $V(r,a)$ depends on a parameter $a \in(a_1,a_2),$ $V(r,a) \le 0,$ and $E(a)$ is any positive eigenvalue, then 
 $\partial V/\partial a \ge 0 \Rightarrow E'(a) \ge  0$ and $\partial V/\partial a \le 0 \Rightarrow E'(a) \le  0.$
\end{abstract}
\vskip0.2in
\maketitle
%%%%%%%%%%%%%%%%%%%%%%%%%%%%%%%%%%%%%%%%%%%%%%%%%%%%%%%%%%%%%%%%%%%%%
\section{Introduction}
%%%%%%%%%%%%%%%%%%%%%%%%%%%%%%%%%%%%%%%%%%%%%%%%%%%%%%%%%%%%%%%%%%%%%
We consider a single particle that is bound by an attractive central potential $V$ in $d$ spatial dimensions. Under the Schr\"odinger theory, for suitable potentials, the Hamiltonian $H = -\Delta + V$ is bounded below and the discrete spectrum may be characterized variationally.  Thus, if $V_1$ and $V_2$ are two potentials which satisfy $V_1 \le V_2,$ then corresponding discrete energy eigenvalues satisfy $E_1 \le E_2.$  In relativistic quantum mechanics, the Hamiltonian is not bounded below and it is not easy to characterize the eigenvalues variationally and thereby to conclude immediately that the energies are ordered if the potentials are.  For node-free ground states, we have been able to prove a comparison theorem for the Dirac equation \cite{halld}.  In the present paper we show in section~2 that similar reasoning yields Theorem~1, a comparison result for the Klein-Gordon equation. Throughout this paper $V$ represents the time component of a four-vector; the scalar potential (a perturbation of the mass) is assumed to be zero.

In section~3 we examine a more specialized comparison problem, namely where an attractive central potential is monotonic in a parameter.  For this class of comparisons, we are able to prove Theorem~2, to the effect that every discrete eigenvalue, in parametric regions where it is positive, is also monotonic in the parameter. Thus, to be more explicit, if $\partial V(r,a)/\partial a \le 0,$ then $E'(a) \le 0$, and, similarly, with the two inequalities reversed.  For example, if we consider a coupling parameter $v >0$ and a potential shape $f(r) \le 0,$ then we  have $V(r,v) = vf(r)$ and $\partial V(r,v)/\partial v = f(r) \le 0.$  In the positive-energy regions, this result confirms the qualitative eigenvalue behaviour of well-known exact analytical solutions for exponential, square-well, and Coulomb potentials found, for example, in Greiner's text book \cite{greiner}. 

A motivation for the present paper came from recent Klein-Gordon studies \cite{barton, hallkg} of cutoff Coulomb potentials, such as the potential $V(r,a) = -\alpha/(r + a)$ of Loudon's classic paper \cite{loudon} on the corresponding non-relativistic problem.  For this potential, and $d = 1$, part of the ground-state Klein-Gordon eigenvalue function $E(a)$ is shown in Fig.~1.
% -----------------------
%  Figure (1)
% -----------------------
\begin{figure}[htbp]
\centering
\includegraphics[width=10 cm]{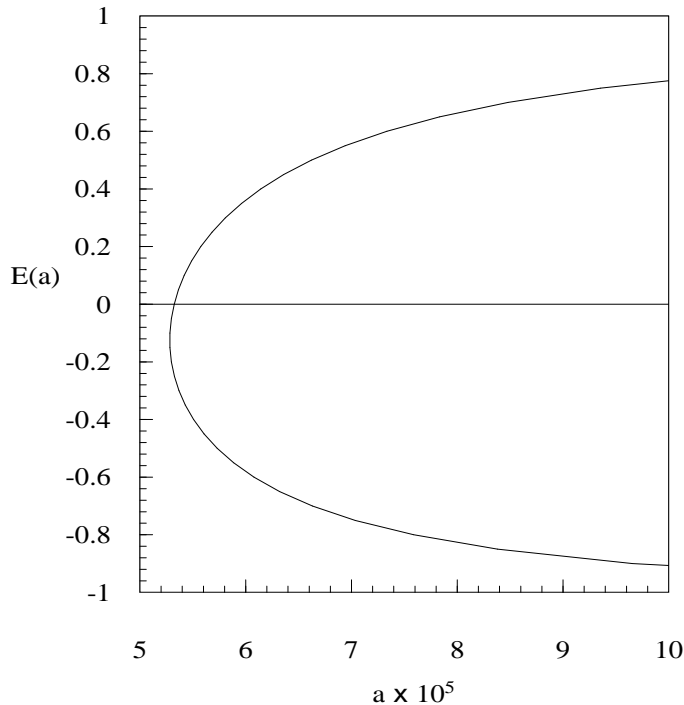}
\caption{Plot of Klein-Gordon ground-state energy $E(a)$ for the potential $V(r,a) = -\alpha/(r+a)$ in dimension $d = 1.$
We use dimensionless units in which $m = c = \hbar = 1,$ and the fine-structure constant $\alpha = 1/137.$}
\label{Fig. 1}
\end{figure}
A question that arises is, how much of this explicit curve can be understood on general grounds?  Since $\partial V(r,a)/\partial a > 0,$ we know {\it a priori} from Theorem~2 that $E'(a) > 0$ if $E(a) > 0.$  This clearly leaves some interesting non-monotonic behaviour still to be explained.  We observe for this problem that the minimum value of $a$ is slightly less than the zero of the two-valued function $E(a):$ thus, whilst it is true that $E'(a) > 0,$ for $E(a) > 0,$ in agreement with Theorem~2, there are both increasing and decreasing regions of the curve for $E(a) < 0.$ More spectral details may be found in Refs. \cite{barton, hallkg} (in which Barton's parametrization is  used for the Coulomb-cutoff spectral curves). 

All the main arguments of the present paper go through independently of the number $d \ge 1$ of spatial dimensions.  In its technical details, the case $d=1$ is special in that it allows even and odd solutions, and the even eigenstates do not necessarily vanish at the origin.  For simplicity we shall present our proofs uniformly for the cases $d > 1,$ claiming  that our reasoning and results apply {\it mutatis mutandis} to the corresponding one-dimensional problem.  In natural units $\hbar = c = 1,$ and the central attractive vector potential $V,$ the Klein-Gordon equation may be written
\begin{equation}\label{eq1}
(-\Delta +  m^2)\Psi = (E -V)^2\Psi.
\end{equation}
If for $d > 1$ we write $\Psi(\mathbf{r}) = r^{-(d-1)/2}\psi(r)Y_{\ell}(\theta_1, \theta_2,\dots \theta_{d-1}),$ where $r = ||\mathbf{r}||,$ and $Y_{\ell}$ is a hyper-spherical harmonic \cite{som} with eigenvalue $\ell(\ell+d-2),$ then Eq.(\ref{eq1}) may be written as the second-order ordinary differential equation:
\begin{equation}\label{eq2}
-\psi''(r) + \frac{Q}{r^2}\psi(r) = \left((E-V(r))^2 -m^2\right)\psi(r), 
\end{equation}
where
\begin{equation}\label{eq3}
Q=\frac{1}{4}(2\ell+d-1)(2\ell+d-3),\quad \ell = 0,1,2,\dots,\ d = 2,3,\dots
\end{equation} 
A useful differential-equation correspondence between all $d > 1$ and $d=3$ can be constructed by writing $Q = L(L+1),$ where
 $L = \ell + (d-3)/2.$ This is particularly useful for extending the range of application of known results for $d = 3:$ one simply replaces $\ell$ by $L$. The radial function $\psi(r)$ in (\ref{eq2}) satisfies $\psi(0) = 0$ and, for bound states, it also satisfies the normalization condition $\int_0^{\infty} \psi^2(r)dr = 1.$
% ------------------------------------------------------
\section{Consequences of a node-free ground state}
We now compare two problems with potentials that are ordered $V_1 \le V_2.$ We suppose that the respective ground states $\{\psi_1(r), \psi_2(r)\}$ in (\ref{eq2}) are node free and we write the corresponding energies as $\{E_1, E_2\}.$ The two eigen equations become:
\begin{eqnarray}
-\psi_1''(r) + \frac{Q}{r^2}\psi_1(r) &=& \left((E_1-V_1(r))^2 -m^2\right)\psi_1(r),\label{eq4}\\ 
-\psi_2''(r) + \frac{Q}{r^2}\psi_2(r) &=& \left((E_2-V_2(r))^2 -m^2\right)\psi_2(r).\label{eq5}
\end{eqnarray}
With these two equations we form the difference (4)$\psi_2(r)$ - (5)$\psi_1(r)$ and integrate this over $[0,\infty)$ to give the following key equation of Theorem~1:
\begin{equation}\label{eq6}
\left(E_2 - E_1\right)\int\limits_0^{\infty}W(r)\psi_1(r)\psi_2(r)dr = \int\limits_0^{\infty}
\left(V_2(r) - V_1(r)\right)W(r)\psi_1(r)\psi_2(r)dr,~
\end{equation}
where
\begin{equation}\label{eq7}
\quad W(r) = E_1 + E_2 - V_1(r) -V_2(r).
\end{equation}
Provided the wave functions are node free, one specific class of definite examples leads immediately to the positivity of $W$ and our first comparison theorem.
\nll{\bf Theorem~1}
\nll{\it If $\psi_1$ and $\psi_2$ are node-free ground states corresponding to positive energies $E_1\ge 0$ and $E_2\ge 0,$
 and $V_1(r)\le V_2(r)\le 0$, then it follows that $E_1 \le E_2.$}
\nll As a simple concrete example we consider in $d=3$ the potentials
\begin{equation}\label{eq8}
-\frac{2}{\left(1+r+\frac{r^2}{2}+\frac{r^3}{6} +\frac{r^4}{24}\right)} = V_1(r)\ \le\  V_2(r) = -2e^{-r}.
\end{equation}
We find numerically for these potentials, which are similar for small $r$, that the Klein-Gordon ground-state eigenvalues are respectively $E_1 =  0.7464$ and $E_2 = 0.7542$, in agreement with Theorem~1. 

A class of examples is generated by negative potentials $V(r) = vf(r),$ where the coupling parameter $v > 0.$  For these problems, if the eigenenergies are positive, and $v_2 > v_1,$ then by Theorem~1, we know $E(v_1) \ge E(v_2).$  However, for potentials depending monotonically on a parameter, Theorem~2, below, is stronger, since it does not need a node assumption, and it applies to every discrete eigenvalue.  A similar result has been derived \cite{halldd} for the Dirac equation in which case the positivity of the energy is not required.

% ------------------------------------------------------
\section{A monotonic family of potentials}
% ------------------------------------------------------
In this section we consider a family of potentials $V(r,a),$ where $a$ is a parameter.  We suppose that discrete eigenvalues exist for the values of the parameter considered. There is no assumption made concerning nodes; in fact, the result of this section is valid for any discrete eigenvalue.   We suppose that there is an  interval $a\in(a_1,a_2)$ in which  $V$ is monotonic in the parameter $a.$  For the convenience of an elementary proof, we shall assume differentiability of $V(r,a)$, $\psi(r,a)$, and $E(a)$ with respect to $a$. We suppose that $E(a)$ is a discrete eigenvalue and that $\psi(r,a)$ is the corresponding radial eigenfunction.  We shall denote partial derivatives of $\psi(r,a)$ with respect to $r$ as $\partial \psi(r,a)/\partial r = \psi'(r,a),$ and partial derivatives with respect to the parameter $a$ by a subscript, thus $\partial\psi(r,a)/\partial a = \psi_a(r,a).$ If we differentiate the normalization integral $\int_0^{\infty}\psi^2(r,a)dr = 1$ partially with respect to the parameter $a,$ we obtain the orthogonality relation $(\psi,\psi_a) = 0.$ Meanwhile, if we define the symmetric operator $K$ as
\begin{equation}\label{eq9}
K = -\frac{\partial^2}{\partial r^2} +\frac{Q}{r^2} + 2EV - V^2,
\end{equation}
then the Klein-Gordon eigenequation (\ref{eq2}) becomes
\begin{equation}\label{eq10}
K\psi = \left(E^2 - m^2\right)\psi
\end{equation}

and we have $\langle K\rangle = (\psi, K\psi) = E^2-m^2.$ Differentiating this last equation partially with respect to $a,$ we obtain
\begin{equation}\label{eq11}
2E(a)E'(a) = (\psi_a,K\psi) + (\psi,K_a\psi) + (\psi,K\psi_a). 
\end{equation}
The symmetry of $K$, and the orthogonality of $\psi$ and $\psi_a$ imply that 
$$(\psi,K\psi_a) = (\psi_a,K\psi) = \left(E^2-m^2\right)(\psi_a,\psi) = 0.$$
Thus, by using the expression
\begin{equation}\label{eq12}
K_a = 2E'(a)V + 2E(a)V_a -2VV_a
\end{equation}
in the middle term on the right-side of (\ref{eq11}), we obtain the key equation for the theorem of this section, namely
\begin{equation}\label{eq13}
E'(a) = \frac{E(a)\langle V_a\rangle - \langle V V_a\rangle}{E(a) -\langle V\rangle}.
\end{equation}

\noindent By making the appropriate assumptions we therefore establish
\nll{\bf Theorem~2}
\nll{\it If for $a\in(a_1,a_2),$ $V(r,a) \le 0,$ and $E(a)\ge 0$ is a corresponding discrete Klein-Gordon eigenvalue, then 
 $\partial V/\partial a \ge 0 \Rightarrow E'(a) \ge  0,$ and $\partial V/\partial a \le 0 \Rightarrow E'(a) \le  0.$}
\nll The coupling as a parameter gives a class of examples.  Thus, for each positive eigenvalue $E(v),$ where $v > 0$ is the coupling to a negative potential
 $V(r) = vf(r),$ we have $E'(v) \le 0.$  The hydrogenic spectrum in $d=3$ and many other exact solutions too the Klein--Gordon spectral problem discussed in Chapter~1 of the book by Greiner \cite{greiner} provide immediate examples of this result.  Another specific illustration is given by the two-parameter cutoff Coulomb potential $V(r) -v/(r +a)$ which generates eigenvalues $E(v,a)$ for positive $v$ and $a$.  We may immediately conclude from Theorem~2 that in any dimension $d$, whenever $E(v,a) > 0,$ we have $\partial E/\partial v \le 0$ and $\partial E/\partial a \ge 0$. These results are consistent with the positive part of the eigenvalue curve shown in Fig.~1 for $d=1.$ Similar results are predicted, and indeed found \cite{hallkg}, for the alternative cutoff Coulomb potential, $V(r) = -v/a, r < a,$  $V(r) = -v/r, r\ge a$ for which we do not have the corresponding exact analytical results at hand. These cutoff Coulomb results in dimension $d=1$ are chosen because of the recent interest in the hydrino problem \cite{dombey,barton}.

% ------------------------------------------------------
\section{Conclusion}
% ------------------------------------------------------
When we consider the Schr\"odinger problem with Hamiltonian $H = \Delta + vf(r)$ and a potential shape $f(r)$ that does not change sign, we know that an energy eigenvalue $E(v)$ is monotonic in the coupling $v > 0.$  It is hard to resist the expectation that something like this ought to be true for the Klein--Gordon equation, even if we are aware of certain counter examples.  In this paper we have established two comparison theorems, one only for node-free ground states, and the other for every discrete eigenvalue of a problem whose potential depends monotonically on a parameter.  The theorems are actually special cases derived from the generating equations (\ref{eq6}) and (\ref{eq13}), and chosen on the basis of simplicity.  As we have made clear with the examples mentioned, these theorems do not by any means characterize all Klein--Gordon spectral ordering effects; but we hope that they provide starting points for further study.

% ------------------------------------------------------   
 \section*{Acknowledgement}
% ------------------------------------------------------
Partial financial support of his research under Grant No.~GP3438 from~the Natural
Sciences and Engineering Research Council of Canada is gratefully acknowledged. 
% -------------------------------------------------------------------------
\nll{\bf References}\smallskip
%%%%%%%%%%%%%%%%%%%%%%%%%%%%%%%%%%%%%%%%%%%%%%%%%%%%%%%%%%%%%%%%%%%%%%%%%%%%%

\end{document}